\documentclass[aps,prd,floatfix,twocolumn,nofootinbib,preprintnumbers,superscriptaddress]{revtex4-2}
\usepackage[utf8]{inputenc}
\usepackage{soul}
\usepackage[table]{xcolor}
\usepackage[compat=1.0.0]{tikz-feynman}
\usepackage{amsfonts}
\usepackage{amsmath}
\usepackage{array} 
\usepackage[markup=nocolor]{changes}
\usepackage{graphicx} 
\usepackage{hyperref}
\usepackage{multirow}
\usepackage{diagbox} 
\usepackage{commath}
\usepackage{pifont}
\usepackage{url}
\usepackage{float} 
\usepackage{xspace}
\usepackage{setspace}
\usepackage{braket}
\usepackage{graphics}
\usepackage{enumitem}
\usepackage{amssymb}
\usepackage{pgfplots}
\graphicspath{{../}}

\hypersetup{pdfstartview=FitV,colorlinks=true,linkcolor=blue,citecolor=blue, filecolor=black,urlcolor=blue}

\newcommand{\be}{\begin{equation}}
\newcommand{\ee}{\end{equation}}
\newcommand{\bea}{\begin{eqnarray}}
\newcommand{\eea}{\end{eqnarray}}

\definecolor{lime}{HTML}{A6CE39}
\DeclareRobustCommand{\orcidicon}{\hspace{-1.8mm}
	\begin{tikzpicture}
	\draw[lime, fill=lime] (0,0) 
	circle [radius=0.16] 
	node[white] at (-0.007,-0.007) {{\fontfamily{qag}\selectfont \tiny \,ID}};
	\draw[white, fill=white] (-0.058,0.095) 
	circle [radius=0.005];
	\end{tikzpicture}
	\hspace{-2.5mm}
}

\foreach \x in {A, ..., Z}{\expandafter\xdef\csname orcid\x\endcsname{\noexpand\href{https://orcid.org/\csname orcidauthor\x\endcsname}
			{\noexpand\orcidicon}}
}
\usepackage{calc}
\newlength{\depthofsumsign}
\makeatletter
\newcommand*{\DivideLengths}[2]{%
  \strip@pt\dimexpr\number\numexpr\number\dimexpr#1\relax*65536/\number\dimexpr#2\relax\relax sp\relax
}
\makeatother

\begin{document}
\title{Fate of Metastable Vacua in the Type-II Two-Higgs-Doublet Model}

\author{Debtosh Chowdhury\orcidA{}} \email{debtoshc@iitk.ac.in}
\affiliation{Department of Physics, Indian Institute of Technology Kanpur, Kanpur 208016, India}
\author{Kirtimaan A. Mohan\orcidD{}} \email{kamohan@msu.edu}
\affiliation{Department of Physics and Astronomy, Michigan State University, 567 Wilson Road, East Lansing, MI-48824, USA}
\author{Poulami Mondal\orcidB{}} \email{poulami.mondal@tifr.res.in}
\affiliation{Department of Theoretical Physics, Tata Institute of Fundamental Research, Mumbai 400005, India}
\author{Subrata Samanta\orcidC{}} \email{samantaphy20@iitk.ac.in}
\affiliation{Department of Physics, Indian Institute of Technology Kanpur, Kanpur 208016, India}

\preprint{TIFR/TH/26-18}
\preprint{MSUHEP-26-006}
\begin{abstract}
\noindent
The scalar potential of the Two-Higgs-Doublet Model (2HDM) can admit multiple non-degenerate vacua due to the presence of the two Higgs doublets unlike the Standard Model (SM). For a physically viable parameter point, one of these vacua must correspond to the physical electroweak (EW) symmetry breaking vacuum with the vacuum expectation value of about $246$ GeV. Given the complex structure of the scalar potential, the physical EW vacuum may be metastable in nature rather than the global minimum of the potential. In this work, we delineate regions of the parameter space in the Type-II 2HDM accommodating multiple extrema of the scalar potential and analyze, in a gauge-independent manner, the stability of the EW vacuum there at the tree level and beyond. A Bayesian global fit of the Type-II 2HDM, including next-to-leading-order (NLO) unitarity constraints and the latest experimental measurements, excludes the parameter space regions that lead to a metastable EW vacuum at both tree and one-loop level, after marginalizing over all other parameters. Additionally, we find that these regions are even more strongly excluded if the NLO perturbativity condition $R_1' < 1$ is imposed.
\end{abstract}

\maketitle
\section{Introduction}\label{sec:intro}
Discovery of the Higgs boson with a mass $\sim 125$ GeV at the Large Hadron Collider (LHC)~\cite{ATLAS:2012yve,CMS:2012qbp} successfully tested the Standard Model (SM) of particle physics. The absolute stability of the Higgs potential within the SM requires the Higgs quartic coupling ($\lambda$) to be positive, i.e., $\lambda>0$, over the entire range of energy scales up to the Planck scale. At the electroweak (EW) scale, the value of $\lambda$ is fixed by the measured Higgs
mass. However, the renormalization group (RG) evolution of $\lambda$ within the SM tends to turn negative at scales between the EW and Planck scales~\cite{Degrassi:2012ry,Alekhin:2012py}. It crosses zero at a scale of approximately $10^{10}$ GeV~\cite{Buttazzo:2013uya,Andreassen:2017rzq,Hiller:2024zjp}, indicating that the EW vacuum is metastable.

To ensure absolute stability of the EW vacuum up to the Planck scale, one needs new particles and their interactions beyond the SM (BSM), see, for instance, Refs.~\cite{Bezrukov:2012sa,Branchina:2013jra,Branchina:2014rva}. One of the simplest and most widely studied extensions of the SM is the Two-Higgs-Doublet Model (2HDM)~\cite{Lee:1973iz,Gunion:2002zf}, (see also Ref.~\cite{Branco:2011iw} for a review) where an additional Higgs doublet is added to the SM Higgs sector. As a result, the 2HDM contains five physical Higgs states: two CP-even neutral scalars, one CP-odd neutral scalar, and a pair of charged scalars. To avoid tree level flavor-changing neutral currents (FCNCs), a $Z_2$ symmetry, under which the two Higgs doublets transform as $\phi_1\to -\phi_1$ and $\phi_2\to \phi_2$, is imposed on the Yukawa sector~\cite{Glashow:1976nt,Paschos:1976ay}. Depending on how the two Higgs doublets couple to fermions, four distinct scenarios arise. In this work, we focus on Type-II 2HDM, in which up-type quarks couple to $\phi_2$, while down-type quarks and charged leptons couple to $\phi_1$. 

In the 2HDM, the presence of additional quartic couplings can alleviate the metastability of the EW vacuum at high scales. Ref.~\cite{Chowdhury:2015yja} showed that there exists  a parameter region in the quartic coupling planes of the 2HDM in which the scalar potential remains stable, i.e. the bounded-from-below (BFB) conditions are satisfied at all scales from the EW scale up to the Planck scale. Additionally, due to the presence of the extra Higgs doublet, the scalar potential of the 2HDM can have several types of minima, including multiple EW breaking minima, as well as the charge breaking (CB) and Charge-Parity (CP) breaking minima. Thus, the scalar potential of the 2HDM can render the EW vacuum metastable at the EW scale. It was shown in Refs.~\cite{Ferreira:2004yd,Barroso:2005sm,Maniatis:2006fs,Nishi:2006tg,Ivanov:2006yq,Ivanov:2007de,Maniatis:2007vn,Barroso:2013awa} that the existence of an EW breaking minimum in the 2HDM scalar potential precludes a \textit{deeper} CB or CP breaking minimum. However, this result is valid only at tree level, as pointed out in Ref.~\cite{Ferreira:2019bij}. In contrast, multiple EW symmetry breaking minima can coexist in
the tree level 2HDM scalar potential~\cite{Ivanov:2006yq,Ivanov:2007de,Barroso:2007rr}. One of these minima must correspond to the physical vacuum with vacuum expectation value (VEV) $v= 246.22~\mathrm{GeV}$, which
we refer to as the \textit{physical EW vacuum}. The remaining neutral extrema correspond to different vacua whose VEVs satisfy $v'^2 \equiv v_1'^2+v_2'^2 \neq (246.22\,\mathrm{GeV})^2$.

This suggests that the physical EW vacuum may not be the absolute minimum of the scalar potential; instead, it may be a metastable state that can later tunnel to the global minimum.
Such an EW vacuum can be phenomenologically viable,  provided its lifetime exceeds the age of the Universe. Refs.~\cite{Barroso:2013awa,Ivanov_2015} developed a formalism and derived analytic conditions for identifying regions of parameter space in which the physical EW vacuum is metastable. Using LHC Run-1 Higgs data, Ref.~\cite{Barroso:2013awa} showed that metastable EW vacua are excluded in the Type-I 2HDM  at tree level, whereas the Higgs data alone do not rule out this possibility in the Type-II 2HDM. This conclusion was further investigated in Refs.~\cite{Barroso:2013ica,Branchina:2018qlf}, which demonstrated that the physical EW vacuum can indeed be metastable in Type-II 2HDM.

However, these studies were carried out primarily using leading-order (LO) theoretical constraints.
The impact of next-to-leading-order (NLO) theoretical constraints, e.g., perturbative unitarity, on the viability of such vacua remains largely unexplored.
The bounds from perturbative unitarity at NLO significantly refine the parameter space in which perturbative calculations can be trusted, as shown in Refs.~\citep{PASSARINO1985231,PhysRevLett.62.1232, Dawson:1989up,PASSARINO199031,PhysRevLett.64.1215,Lendvai:1991gd,Durand:1992wb} for the SM and in Refs.~\cite{Chowdhury:2015yja,Grinstein:2015rtl,Cacchio:2016qyh} for the 2HDM. In particular, Ref.~\cite{Grinstein:2015rtl} showed that the NLO unitarity conditions can be satisfied for quartic couplings as large as $\lambda_1\sim 10$ within the 2HDM. However, perturbation theory is not reliable for such large couplings~\cite{Chowdhury:2015yja}, and therefore Ref.~\cite{Grinstein:2015rtl} further advocated a perturbativity criterion, i.e., the ratio of the NLO to LO partial-wave amplitudes to remain below unity, to delineate parameter regions where perturbative calculations are reliable. According to this criterion, perturbativity is violated when $\lambda_1 \gtrsim 5$~\cite{Grinstein:2015rtl,Cacchio:2016qyh}. Since NLO unitarity and perturbativity requirements impose one of the most stringent constraints on the scalar quartic couplings~\cite{Chowdhury:2015yja,Grinstein:2015rtl,Cacchio:2016qyh}, it is important to incorporate these constraints into the study of the  scalar potential vacuum structure and accurately analyze the nature of the physical EW vacuum. Such an analysis must also incorporate the latest experimental results, including Higgs signal strength measurements, flavor observables, and EW precision constraints from the oblique parameters, all of which can significantly constrain the viable parameter space. Ref.~\cite{Chakraborty:2015raa} also investigated the vacuum structure of the Type-II 2HDM and found that the physical EW minimum remains the global minimum at tree level after imposing the theoretical and experimental constraints considered in their analysis. Their study was, however, restricted to a relatively small range of quartic couplings ($|\lambda_i| \leq 1$). In contrast, the present work explores a substantially broader region of the parameter space through a comprehensive global analysis.

In addition, one-loop Coleman-Weinberg (CW) corrections can modify the vacuum structure of the scalar potential: a minimum that is only local at tree level may become the global minimum at one loop, and vice versa. Thus, the radiative corrections could render the physical EW vacuum metastable~\cite{Ferreira:2015pfi, Basler:2018cwe, Hollik:2018wrr, Ferreira:2019bij, Basler:2024aaf,Chakraborty:2015raa}. Therefore, the stability of the physical EW vacuum must be reassessed once these corrections are included. As pointed out in Refs.~\cite{Nielsen:1975fs,Fukuda:1975di,Patel:2011th,Andreassen:2014eha,Ekstedt:2018ftj,Balui:2025kat}, the CW corrected potential is gauge dependent, but its value at an extremum is gauge invariant. We focus on the possibility that the physical EW vacuum becomes metastable due to the presence of other neutral extrema but not due to the presence of charge and CP breaking extrema.\footnote{One-loop corrections can also generate CB or CP breaking minima that have no tree level counterpart~\cite{Ferreira:2019bij}; locating these requires a direct minimization of the one-loop effective potential rather than the $\hslash$ expansion. We discuss this further in Sec.~\ref{sec:eff2hdm}.}
We assess metastability by comparing the values of the potential at these extrema; since this quantity is gauge invariant, we are free to work in any gauge and, for
convenience, choose the Landau gauge. However, the one-loop CW corrected potential (in the $\overline{\textrm{MS}}$ scheme) also depends explicitly and implicitly on the renormalization scale $\mu$  through the renormalization group running of the quartic couplings. Ref.~\cite{Chakraborty:2015raa} adopts a prescription in which the value of $\mu$ is chosen such that the physical EW vacuum remains at its tree level position even after including one-loop corrections. In this work, we instead fix the renormalization scale to $\mu=v$ and perform the vacuum analysis using the $\hslash$ expansion method~\cite{Patel:2011th}. The extrema of the scalar potential and the depths of these extrema are therefore determined from the one-loop CW corrected potential, order by order in $\hslash$, in a gauge-invariant way~\cite{Andreassen:2014eha,Ekstedt:2018ftj}.

All these subtleties in the tree level and the CW corrected scalar potential motivate a comprehensive reassessment of vacuum metastability in the Type-II 2HDM. In the present study, we investigate whether a metastable physical EW vacuum can persist once improved theoretical constraints and the latest experimental data are taken into account, while considering the scalar potential at tree level and beyond. To this end, we combine the following ingredients that, to the best of our knowledge, have not previously been incorporated simultaneously in an analysis of vacuum metastability in the Type-II 2HDM: \begin{enumerate}
\renewcommand{\labelenumi}{(\roman{enumi})}
\item we incorporate improved theoretical constraints, namely NLO unitarity, perturbativity, and bounded-from-below (BFB) conditions; 
\item we perform a Bayesian global fit including the latest flavor, EW precision, and Higgs data, together with the improved theoretical constraints. Within the viable parameter space, we examine the vacuum structure of the scalar potential;
\item we employ the gauge-invariant $\hslash$ expansion method~\cite{Patel:2011th} to study the stability of the physical EW vacuum beyond tree level.\footnote{We specifically look at the impact of loop corrections to the vacua that are not CB or CP breaking, and leave the analysis of CB or CP breaking minima for future work.}
\end{enumerate}

The rest of the paper is structured as follows. In Section~\ref{sec:2hdm}, we briefly review the Type-II 2HDM. In Section~\ref{sec:extrema}, we discuss the vacuum structure of the tree level scalar potential and outline the conditions for the existence of multiple extrema in the field space. Section~\ref{sec:unitarity} summarizes the NLO unitarity constraints and the perturbativity criterion employed in our analysis. In Section~\ref{sec:eff2hdm}, we review the one-loop CW corrected potential for the 2HDM and discuss its gauge dependence.  The theoretical and experimental constraints, together with the parameter-sampling procedure, are presented in Section~\ref{sec:constraints}. Our results are presented in Section~\ref{sec:results}, and our conclusions are given in Section~\ref{sec:conclusions}. Additional figures and supplemental results are provided in Appendix~\ref{app:supp_plots}.

\section{Two-Higgs-Doublet Model}\label{sec:2hdm}
By extending the SM by an additional Higgs doublet, the 2HDM~\cite{Lee:1973iz,Gunion:2002zf} gives rise to a considerably richer scalar sector and phenomenology.  The most general CP conserving 2HDM scalar potential contains ten real parameters.  To suppress tree level FCNCs, a discrete $Z_2$ symmetry is imposed on the Yukawa sector. Depending on the $\mathbb{Z}_2$ charge assignments of the right-handed quarks and charged leptons, four distinct Yukawa structures arise, commonly referred to as the Type-I, Type-II, Type-X (lepton-specific), and Type-Y (flipped) 2HDMs~\cite{Branco:2011iw}. In this work, we focus on the Type-II 2HDM, which also corresponds to the scalar sector of the minimal supersymmetric Standard Model. The scalar potential of the CP conserving Type-II 2HDM with a softly broken $\mathbb{Z}_2$ symmetry contains eight real parameters and
is given by 
    \begin{align}\label{eq:pot-tree}
        V= \;& m_{11}^2|\phi_1|^2+m_{22}^2|\phi_2|^2-m_{12}^2(\phi_1^\dagger\phi_2+\mathrm{h.c.})\nonumber\\
        &+\frac{1}{2}\lambda_1|\phi_1|^4+\frac{1}{2}\lambda_2|\phi_2|^4+\lambda_3|\phi_1|^2|\phi_2|^2\nonumber\\
        &+\lambda_4|\phi_1^\dagger\phi_2|^2+\frac{1}{2}\lambda_5[(\phi_1^\dagger\phi_2)^2+\mathrm{h.c.}]\,,
    \end{align}
    where 
    \begin{align}
        \label{eq:2HDM_fields}
        \phi_a=
        \begin{pmatrix}
        \chi_a^+  \\
\frac{1}{\sqrt{2}}(h_a + i \eta_a)
        \end{pmatrix}\,, \quad a=1,2\,.
    \end{align}
    Here $m_{11}^2$, $m_{22}^2$, $m_{12}^2$, and $\lambda_{1,2,3,4,5}$ are all taken to be real. At the minimum of the scalar potential,  the neutral components $h_1$ and $h_2$ acquire VEVs, $v_1$ and $v_2$, respectively. Apart from the three Goldstone bosons, the scalar spectrum consists of two CP-even states
$h$ and $H$, a CP-odd state $A$, and a pair of charged states $H^\pm$. The analytic expressions of the mass-squared eigenvalues of these states are given by
\begin{align}
\label{eq:masses}
    m_{H,h}^2&=\frac{1}{2}\left[M_{11}^2+M_{22}^2\pm \sqrt{(M_{11}^2-M_{22}^2)^2+4(M_{12}^2)^2}\right],\nonumber\\
        m_{H^\pm}^2&=\left[\frac{m_{12}^2}{v_1v_2}-\frac{1}{2}(\lambda_4+\lambda_5)\right]v^2,\nonumber\\
         m_{A}^2&=\left(\frac{m_{12}^2}{v_1v_2}-\lambda_5\right)v^2,
\end{align}
where $v^2=v_1^2+v_2^2$, and 
\begin{align}
\label{eq:M11-M22}
M_{11}^2=& m_{12}^2\frac{v_2}{v_1}+\lambda_1 v_1^2\,,\nonumber\\
M_{22}^2=& m_{12}^2\frac{v_1}{v_2}+\lambda_2 v_2^2\,,\nonumber\\
M_{12}^2=& -m_{12}^2+(\lambda_3+\lambda_4+\lambda_5) v_1 v_2\,.
\end{align}

Furthermore, for the Type-II 2HDM considering only the third generation of fermions, the $Z_2$-symmetric Yukawa Lagrangian is given by~\cite{Branco:2011iw},
\begin{equation}
    \mathcal{L}_Y=-y_t\bar{Q}_Li\sigma_2\phi_2^*t_R-y_b\bar{Q}_L\phi_1b_R-y_\tau\bar{L}_L\phi_1\tau_R +\text{h.c.}\,,
\end{equation}
where $y_t=y_t^{\text{SM}}/\sin\beta$,  $y_f=y_f^{\text{SM}}/\cos\beta$ for $f=b,\tau$, and the mixing angle $\beta\in[0,\pi/2]$ is defined by  $\tan\beta=v_2/v_1$.

In the SM, there exists only one type of vacuum, preserving CP and $U(1)_{\text{EM}}$ but spontaneously breaking $SU(2)_L\times U(1)_Y$ symmetry, whereas in the 2HDM, there are three types: (i) EW breaking vacua as in the SM ($V_N$), (ii) vacua which spontaneously break CP ($V_{CP}$), (iii) charge breaking vacua ($V_{CB}$). If $m_{H^\pm}^2>0$ at the EW breaking vacuum, the difference in potential depths satisfies the condition $V_{\text{CB}}-V_N>0$, so the CB stationary point lies above the EW vacuum and is necessarily a saddle point; the EW vacuum is thus stable against CB vacuum~\cite{Ferreira:2004yd,Barroso:2005sm}.
Likewise, for $m_A^2>0$ at the EW breaking vacuum, one finds $V_{\text{CP}}-V_N>0$, ensuring stability against CP breaking minima~\cite{Ferreira:2004yd,Barroso:2005sm}. Hence, minima of different types cannot coexist in the 2HDM: whenever an EW breaking minimum exists, any CB or CP breaking stationary point is necessarily a saddle point lying above it~\cite{Ferreira:2004yd,Barroso:2005sm,Barroso:2013awa}.\footnote{It is worth noting that this is true only when we consider the scalar potential at the tree level~\cite{Ferreira:2019bij}.} This does not, however, preclude a second, deeper neutral EW breaking minimum $N'$, the possibility we examine below and which is referred to as ``\textit{panic vacuum}'' in Refs.~\cite{Barroso:2012mj,Barroso:2013awa}.

As noted above, in addition to the physical EW minimum $N$, characterized by
$v^2 = v_1^2 + v_2^2 = (246.22\,\text{GeV})^2$, a second neutral extremum $N'$ with vacuum expectation values $v_1',v_2'$, and
$v_1'^2 + v_2'^2 \neq (246.22\,\text{GeV})^2$
may also exist.
The difference in potential depth between $N$ and $N'$ is given by~\cite{Barroso:2013awa},
    \begin{equation}
        V_{N'}-V_{N}=\frac{m_{12}^2}{4v_1v_2}\left(1-\frac{v_1v_2}{v_1'v_2'}\right)(v_1v_2'-v_2v_1')^2\,.
    \end{equation}
          If $m_{12}^2=0$, the two EW breaking minima $N$ and $N'$ become degenerate. This corresponds to an exact $Z_2$ symmetry ($\phi_1\to -\phi_1$, $\phi_2\to\phi_2$) of the scalar potential. Importantly, quantum corrections can alter the relative depths of these extrema and thereby modify the vacuum structure of the theory~\cite{Chakraborty:2015raa,Ferreira:2015pfi}. In particular, the ordering of two neutral minima can be reversed between the tree and one-loop levels: a configuration with $V_N - V_{N'} < 0$ at tree level, so that the physical EW vacuum $N$ is the global minimum, can instead have $V_N - V_{N'} > 0$ once one-loop corrections are included, rendering the EW vacuum metastable. We focus on precisely this possibility and determine whether, given current experimental and theoretical constraints, such metastable EW vacua are allowed or disfavored. To this end, in the following sections, we first review the conditions for the existence of tree level stationary points before describing our setup for computing the one-loop corrections.

\section{Stationary Points of the tree level Potential}\label{sec:extrema}
The tree level vacuum structure of the 2HDM has been studied extensively~\cite{Ferreira:2004yd,Barroso:2005sm,Maniatis:2006fs,Nishi:2006tg,Ivanov:2006yq,Ivanov:2007de,Barroso:2007rr,Maniatis:2007vn,Barroso:2013awa}; here we recall the results relevant to our analysis. Following the notation of Refs.~\cite{Ivanov:2006yq,Ivanov:2007de}, we introduce the following gauge-invariant bilinears in terms of hyperspinor $\Phi=\begin{pmatrix} \phi_1 & \phi_2 \end{pmatrix}^T$,
\begin{equation}
    r_0=\Phi^\dagger\Phi\,,\qquad r_a=\Phi^\dagger\sigma_a\Phi\,,
\end{equation}
where $ \phi_1 , \phi_2 $ are the two Higgs doublets, and $\sigma_a\; (a=1,2,3) $ are the Pauli matrices. The scalar potential in this notation can be written in a compact form,
\begin{equation}\label{eq:Vorbit}
 V=-M_\mu r^\mu+\frac{1}{2}\Lambda_{\mu \nu}r^\mu r^\nu \,,  
\end{equation}
with 
\begingroup
\small
\begin{align}
    M_\mu&=\begin{pmatrix}-\dfrac{m_{11}^2+m_{22}^2}{2},& m_{12}^2,&0,&\dfrac{m_{22}^2-m_{11}^2}{2}\end{pmatrix},\\[9pt]
   \Lambda_{\mu\nu}&=\dfrac{1}{2}\begin{pmatrix}
        \dfrac{\lambda_1+\lambda_2}{2}+\lambda_3 &0 &0 & \dfrac{\lambda_1-\lambda_2}{2}\\
        0& \lambda_4+\lambda_5 & 0& 0\\
        0 &0& \lambda_4-\lambda_5&0\\
        \dfrac{\lambda_1-\lambda_2}{2} &0&0& \dfrac{\lambda_1+\lambda_2}{2}-\lambda_3
    \end{pmatrix}.\nonumber
\end{align}
\endgroup
The scalar potential can then be analyzed in terms of the four-vector $r_\mu$ ($\mu = 0,1,2,3$), restricted to the physical orbit space defined by $r_0 \ge 0$ and $r_\mu r^\mu \ge 0$. These constraints ensure that $r_\mu$ corresponds to gauge-invariant bilinears of the scalar fields and allow a geometric classification of extrema in field space~\cite{Ivanov:2006yq, Ivanov:2007de}. For $r_0=0$, the only solution is the trivial configuration
$\phi_1=\phi_2=0$, for which $V=0$. On the boundary of the orbit space, defined by $r_\mu r^\mu = 0$, the stationary points correspond to neutral extrema. Minimization of the potential under this constraint leads to
\begin{equation}\label{eq:neutralSP}
    \Lambda_{\mu\nu}r^\nu-M_\mu=\zeta r_\mu\,,
\end{equation}
where $\zeta$ is the Lagrange multiplier.
In the basis where $\Lambda_{\mu\nu}$ is diagonal, Eq.~(\ref{eq:neutralSP}) becomes 
\begin{equation}\label{eq:eigenBasisSPE}
    \Big(\Lambda_0-\zeta\Big)\hat{r}_0=\hat{M}_0\,,\qquad \Big(\Lambda_a-\zeta\Big)\hat{r}_a=\hat{M}_a\,,
\end{equation}
where the eigenvalues of $\Lambda_{\mu\nu}$ are given by
\begin{align}\label{eq:LAMBDA-eigen}
    \Lambda_0&=\frac{1}{2}\Big(\lambda_3+\sqrt{\lambda_1\lambda_2}\Big)\,,&\Lambda_1&=-\frac{1}{2}\Big(\lambda_4+\lambda_5\Big)\,, \nonumber\\[3pt]\Lambda_2&=-\frac{1}{2}\Big(\lambda_4-\lambda_5\Big)\,,&\Lambda_3&=\frac{1}{2}\Big(\lambda_3-\sqrt{\lambda_1\lambda_2}\Big)\,,
\end{align}
and the components of $M_\mu$ in this basis are
\begin{align}
    \hat{M}_0&=-\frac{1}{2k}\Big(m_{11}^2+k^2m_{22}^2\Big)\,,& \hat{M}_1&=m_{12}^2\,,\nonumber\\
    \hat{M}_3&=\frac{1}{2k}\Big(k^2m_{22}^2-m_{11}^2\Big)\,,&\hat{M}_2&=0\,,
\end{align}
with $k^2=\sqrt{\lambda_1/\lambda_2}$\,. As shown in Ref.~\cite{Ivanov:2007de}, the existence and multiplicity of such extrema depend on the invariant $\hat M_\mu \hat M^\mu$. 
For $\hat M_0 < 0$ and $\hat M_\mu \hat M^\mu \ge 0$, no non-trivial neutral solution exists. In contrast, for $\hat M_0 < 0$ and
$\hat M_\mu \hat M^\mu < 0$, there exists a unique
neutral stationary point corresponding to the physical
EW minimum. Consequently, $\hat M_0 > 0$ is a necessary condition for the scalar potential to admit more than one neutral stationary point.
For $r_\mu r^\mu>0$, all stationary points correspond to CB vacua, which satisfy
\begin{equation}\label{eq:CB-vacua}
    \Lambda_{\mu\nu}r^\nu=M_\mu\,.
\end{equation}
In the $\Lambda_{\mu\nu}$-diagonal basis, Eq.~(\ref{eq:CB-vacua}) reduces to
\begin{equation}\label{eq:mutlextrcond}
    \Lambda_0 \hat{r}_0=\hat{M}_0\,,\qquad \Lambda_a \hat{r}_a=\hat{M}_a\,.
\end{equation}
The sign of $\hat M_0$ in the above relation is therefore controlled by $\operatorname{sgn}(\Lambda_0)$. Requiring the scalar potential in Eq.~(\ref{eq:pot-tree}) to be bounded from below imposes the following BFB conditions on the quartic couplings~\cite{Deshpande:1977rw},
\begin{align}
\lambda_1 &> 0 , \quad \lambda_2 > 0 ,  \notag  \\ 
\lambda_3 &> -\sqrt{\lambda_1 \lambda_2} , \quad 
\lambda_3 + \lambda_4 - |\lambda_5| > -\sqrt{\lambda_1 \lambda_2} , \label{eq:BFB2}
\end{align}
which in particular guarantee that $\Lambda_0>0$. Since $\hat r_0>0$, the first condition in Eq.~\eqref{eq:mutlextrcond} implies that $\hat M_0>0$. Hence, $\hat M_0>0$ is a necessary condition for the tree level scalar potential to admit multiple stationary points, whether neutral or charge breaking. In the following, we refer to this condition (i.e., $\hat{M}_0>0$) as the \textit{multiple extrema condition}. At the EW breaking vacuum, this condition can be recast in terms of physical parameters as
\begin{equation}
m_{H^\pm}^2-\Lambda_0 v^2<0\,,
\label{eq:upperbound_mH+}
\end{equation}
or, equivalently,
\begin{equation}
m_{12}^2+(\Lambda_1-\Lambda_0)v_1v_2<0\,,
\label{eq:upperbound_m12sq}
\end{equation}
where $\Lambda_0$ and $\Lambda_1$ are given in Eq.~(\ref{eq:LAMBDA-eigen}), and $v^2=v_1^2+v_2^2$. Eq.~(\ref{eq:upperbound_mH+}) and Eq.~(\ref{eq:upperbound_m12sq}) imply upper bounds on $m_{H^\pm}$ or $m_{12}^2$ in terms of the quartic couplings. Since perturbative unitarity constrains the quartic couplings, and hence $\Lambda_0$ and $\Lambda_1$, it follows that unitarity bounds translate into corresponding upper limits on $m_{H^\pm}$ and $m_{12}^2$. In particular, NLO perturbative unitarity constraints are more restrictive, and as a result lead to stronger upper bounds on $m_{H^\pm}$ and $m_{12}^2$. We review these LO and NLO perturbative unitarity constraints in the following section.

\section{Next-to-Leading-Order Perturbative Unitarity}\label{sec:unitarity}
Requiring the $S$ matrix to be unitary bounds the quartic couplings through the amplitudes for all $2\to2$ scatterings among the scalar and would-be-Goldstone components of the two Higgs doublets. By the optical theorem, the spin-zero partial wave $a_0$ must satisfy $\mathrm{Im}(a_0)\ge|a_0|^2$, i.e., it must lie on or within the Argand circle of radius $1/2$ centered at $i/2$ in the complex plane. In the 2HDM these $a_0$ are the eigenvalues of the matrix of such amplitudes~\cite{Ginzburg:2005dt}.

At tree level the eigenvalues are real and the condition reduces to $|\mathrm{Re}\,a_0^{\mathrm{LO}}|\le\tfrac12$~\cite{Ginzburg:2005dt}, which we refer to as the LO unitarity condition. Beyond tree level the amplitude develops an imaginary part, $a_0=a_0^{\mathrm{LO}}+a_0^{\mathrm{NLO}}$, and the full condition $|a_0-\tfrac{i}{2}|\le\tfrac12$ must be imposed at a chosen renormalization scale, which we refer to as the NLO unitarity condition. Since a coupling can satisfy the LO bound while violating the NLO one, the NLO unitarity condition can be significantly more restrictive.

Imposing NLO unitarity alone, however, does not guarantee that the perturbative expansion is reliable: the NLO unitarity bound can still be satisfied for quartic couplings as large as $\lambda_1\sim10$~\cite{Grinstein:2015rtl}, where the one-loop contribution to the amplitude becomes comparable to the tree level one. As a measure of this, Ref.~\cite{Grinstein:2015rtl} proposed, as a minimal requirement for perturbation theory to hold, that the one-loop contribution be smaller than the tree level one, $R_1\equiv|a_0^{\mathrm{NLO}}/(a_0^{\mathrm{LO}}+a_0^{\mathrm{NLO}})|<1$ or  $R_1'\equiv|a_0^{\mathrm{NLO}}/a_0^{\mathrm{LO}}|<1$. In contrast to the unitarity bounds, which follow rigorously from the optical theorem, this perturbativity condition is best understood as a practical guide to the domain of validity of perturbation theory rather than a strict bound, and its precise form (the threshold value and the use of $R_1'$ rather than $R_1$) is to some extent a matter of convention. We therefore regard it as a weaker, supplementary criterion and investigate its impact separately.

\section{One-loop Coleman-Weinberg corrected potential}\label{sec:eff2hdm}
Here, we briefly review the CW corrected potential at the one-loop level and discuss its gauge dependence. In quantum field theory, the tree level scalar potential does not represent the full effective potential, as it receives contributions from quantum corrections. Consequently, loop effects can modify the vacuum structure of the 2HDM. The CW corrected potential of the 2HDM has been studied up to one-loop order in Refs.~\cite{Lee:2012jn,Cao:2022rgh} and extended to two-loop order in Ref.~\cite{Eichten:2022vys}. Although the effective potential is gauge dependent, Refs.~\cite{Nielsen:1975fs,Fukuda:1975di,Patel:2011th,Andreassen:2014eha} showed that the value of the CW corrected potential at its extrema is gauge invariant order by order in perturbation theory, while the field values remain gauge dependent. In the Landau gauge ($\xi=0$), the CW corrected potential for the softly broken $Z_2$-symmetric 2HDM can be written as (at the renormalization scale $\mu$)
\begin{equation}
    V_\text{eff}=V^{(0)}+V^{(1)}+ \cdots\,,
\end{equation}
where $V^{(0)}$ is the tree level potential given in Eq.~(\ref{eq:pot-tree}) and 
\begin{align}
        V^{(1)}=\;& \frac{1}{64\pi^2}\sum_{i=\{h,H,A,H^+,H^-\}} M_i^4\left[\log\left(\frac{M_i^2}{\mu^2}\right)-\frac{3}{2}\right]\nonumber\\
        +&\frac{3}{64\pi^2}\sum_{i=\{Z,W^+,W^-\}} M_i^4\left[\log\left(\frac{M_i^2}{\mu^2}\right)-\frac{5}{6}\right]\nonumber\\
        -&\frac{12}{64\pi^2} M_t^4\left[\log\left(\frac{M_t^2}{\mu^2}\right)-\frac{3}{2}\right]\,,
\end{align}
where we have neglected contributions coming from fermions lighter than the top quark, and the Goldstone boson contributions vanish in the $\xi=0$ gauge since their masses are proportional to the gauge-fixing parameter $\xi$. The remaining field-dependent mass-squared parameters $\{M_i^2\}$ are obtained from the tree level potential evaluated at an extremum. As mentioned above, the field values at the extrema are gauge dependent. The approach of Ref.~\cite{Patel:2011th}, known as the $\hslash$ expansion method, determines the depth of the CW corrected potential at its extrema in a gauge-invariant manner. 

Quantum corrections shift the positions of the extrema and can change the nature of the vacuum, so the vacuum structure must be re-examined beyond the tree level. The $\hslash$ expansion~\cite{Patel:2011th} does this in a gauge-invariant way, by evaluating the potential at each tree level extrema and correcting the depth order by order. At a genuine minimum, the tree level squared masses are positive, which renders the effective potential real-valued. At a saddle point, however, such as a CB or CP breaking vacuum, at least one tree level squared mass is negative, according to the tree level theorems of Refs.~\cite{Ferreira:2004yd,Barroso:2005sm,Barroso:2013awa}. As a result, the one-loop correction, due to the presence of logarithms, turns imaginary. Such cases can still be handled, either by carrying the $\hslash$ expansion to two-loop order and resumming the problematic masses, or by minimizing the full one-loop potential numerically~\cite{Ferreira:2019bij}.\footnote{Taking the direct-minimization route, Ref.~\cite{Ferreira:2019bij} found that quantum corrections can produce what is forbidden at tree level: a CB minimum coexisting with the EW minimum, which can even be deeper, though numerically rare. As in the analogous neutral-minimum case~\cite{Ferreira:2015pfi}, such loop-induced swaps require the relevant tree level extrema to be nearly degenerate, so they reflect small perturbations rather than a breakdown of perturbation theory.} The latter is simpler, but it brings in incomplete higher-order terms that depend on the gauge, so its depths are gauge invariant only up to small two-loop effects~\cite{Andreassen:2014eha,Ekstedt:2018ftj}.

In this work, we compare only neutral, CP-conserving EW vacua, which are ordinary minima already at tree level. For these, the one-loop scalar potential is real, and none of the above complications arise, so the $\hslash$ expansion method provides a true, gauge-invariant comparison of the depths. We therefore use it throughout in our analysis. The procedure consists of the following steps:
\begin{itemize}
    \item Determine the vacuum configurations ($\vec{\phi}^{(0)}_{\text{vac}}$) that extremize the tree level scalar potential. 
    \item Incorporate perturbative corrections to these extrema order by order~\cite{Patel:2011th}. At one-loop order, the effective potential is given by,
    \begin{equation}
    V_\text{eff}(\vec{\phi}_{\text{vac}})=V^{(0)}(\vec{\phi}^{(0)}_{\text{vac}})+ \hslash V^{(1)}(\vec{\phi}^{(0)}_{\text{vac}})\,.
        \label{eq:hbar}
    \end{equation}
\end{itemize}
In our one-loop analysis, for each benchmark point, we compute $V_{\rm eff}(\vec{\phi}_{\rm vac})$ at all EW symmetry breaking extrema and identify the global minimum. As discussed in Ref.~\cite{Andreassen:2014eha}, the value of the effective potential at its minimum is independent of the renormalization scale $\mu$. For convenience, we choose $\mu=v$ throughout the analysis and evaluate the one-loop effective potential.\footnote{We note that Ref.~\cite{Chakraborty:2015raa} also studied the one-loop effective potential and evaluated the depths of the extrema by tuning the renormalization scale $\mu$ such that the physical EW minimum remains at its tree level position even after the inclusion of one-loop CW corrections. Consequently, the renormalization scale $\mu$ is benchmark dependent and must be determined separately for each benchmark point.} 

\section{Sampling Procedure and Constraints}\label{sec:constraints}
The 2HDM parameter space is sampled using the open-source code \texttt{HEPfit}~\cite{DeBlas:2019ehy}, which employs a Bayesian Markov Chain Monte Carlo (MCMC) sampling method based on the Metropolis–Hastings algorithm implemented through the \texttt{Bayesian Analysis Toolkit (BAT)} library~\cite{Caldwell:2008fw}. Theoretical constraints and experimental bounds are incorporated through the corresponding likelihood functions. We consider the lightest CP-even scalar $h$ with fixed
mass $m_h = 125.09$ GeV~\cite{ATLAS:2015yey}, and the Higgs VEV, $v = 246.22$ GeV. The remaining input parameters and the corresponding priors are given below:
\begin{align}
\label{tab:priors}
    -0.3&\leq \log_{10}(\tan\beta)\leq 1.7\,,\nonumber\\
    0& \leq \beta-\alpha\leq \pi\,,\nonumber\\
    -0.2\;\text{TeV}^2& \leq m_{12}^2\leq 0.8\;\text{TeV}^2\,,\nonumber\\
    (0.13\;\text{TeV})^2& \leq m_{H}^2,m_A^2,m_{H^\pm}^2\leq (1.1\;\text{TeV})^2\,,
\end{align}
where $\alpha$ is the mixing angle between the CP-even scalars. 
All sampled points satisfy the two-loop renormalization group improved BFB and perturbativity conditions up to a scale of  
1 TeV.\footnote{The complete list of two-loop beta functions in the 2HDM is given in~\cite{Chowdhury:2015yja}.} In addition, we impose the perturbative unitarity constraints defined in Section~\ref{sec:unitarity}:
\begin{itemize}
    \item the LO unitarity condition;
    \item the NLO unitarity condition, evaluated at a scale of 1 TeV; and 
    \item the perturbativity criterion $R_1'<1$, whose impact we examine separately.
\end{itemize}

For a detailed discussion of the NLO unitarity constraints in the 2HDM and their implementation, we refer to Refs.~\cite{Grinstein:2015rtl,Cacchio:2016qyh}. The experimental constraints included in our analysis are as follows.
\begin{itemize}
    \item the Peskin-Takeuchi parameters $S$, $T$, and $U$~\cite{Haber:1993wf};
    \item the Higgs signal strengths~\cite{Chowdhury:2024mfu}; and 
    \item the $B_s$-meson mass difference ($\Delta m_{B_s}$)~\cite{HeavyFlavorAveragingGroupHFLAV:2024ctg} and the branching ratio $\mathcal{B}(\bar{B }\to X_s \gamma)$~\cite{HeavyFlavorAveragingGroupHFLAV:2024ctg}.
\end{itemize}
The experimental inputs for $S$, $T$, and $U$, together with their correlation coefficients, are taken from the latest EW precision fit performed with \texttt{HEPfit}~\cite{deBlas:2017wmn}, summarized in Table~\ref{tab:expt-data}. 
The Higgs signal strength measurements used in this analysis are based on the latest ATLAS and CMS Run-2 data at a center-of-mass energy of 13 TeV and are listed in Tables 4 and 5 of Ref.~\cite{Chowdhury:2024mfu}. Apart from that, flavor observables provide complementary constraints on the 2HDM parameter space. A comprehensive study of flavor observables and their corresponding constraints on the 2HDM parameter space was performed in Ref.~\cite{Enomoto:2015wbn}, which showed that in the Type-II 2HDM the branching ratio $\mathcal{B}(\bar{B} \to X_s \gamma)$ yields the dominant lower bound on $m_{H^\pm}$, while the $B_s$-meson mass difference $\Delta m_{B_s}$ provides the most stringent constraint on $\tan\beta$ in the low-$\tan\beta$ regime. Given our fit prior $\log_{10}(\tan\beta)\leq 1.7$ (equivalent to $\tan\beta\lesssim 50$), we find that the region with $\tan\beta<4$ is allowed at the $95.4\%$ probability level by the Higgs signal-strength measurements and the constraints from the $S$, $T$, and $U$ parameters (see Figure~\ref{fig:A2} in Appendix~\ref{app:supp_plots}). The flavor observables $\mathcal{B}(\bar{B}\to X_s \gamma)$ and $\Delta m_{B_s}$ are therefore particularly relevant for the parameter region of interest.  The current experimental values of these flavor observables~\cite{HeavyFlavorAveragingGroupHFLAV:2024ctg} are listed in Table~\ref{tab:expt-data2}.

Given the fact that the latest experimental measurement of the muon anomalous magnetic moment~\cite{CMD-3:2023alj,CMD-3:2023rfe} is in agreement with the SM prediction~\cite{Boccaletti:2024guq}, including this observable in our global fit would further constrain the allowed parameter space of the Type-II 2HDM and thereby strengthen the conclusion drawn from $\mathcal{B}(\bar{B}\to X_s\gamma)$.

\begin{table}[ht]
\centering
\renewcommand{\arraystretch}{1.2}
\begin{tabular}{c|c |c c c c}
\hline\hline
Observable  & Value & \multicolumn{3}{c}{Correlation}  \\
\hline
$S$   & $0.09 \pm  0.10  $ & 1.00 & & \\
$T$   & $ 0.11 \pm 0.12 $ & 0.86  & 1.00 & \\
$U$   & $-0.01 \pm 0.09$ & $-0.56$ & $-0.84$ & 1.00\\
\hline\hline
\end{tabular}
\caption{Best fit values of the $S,T,$ and $U$ parameters and their correlation~\cite{deBlas:2017wmn}.}
\label{tab:expt-data}
\end{table}

\begin{table}[ht]
\centering
\renewcommand{\arraystretch}{1.2}
\begin{tabular}{c|c |c  }
\hline\hline
Observable  & Value & Source \\
\hline
$\Delta m_{B_s}$   & $17.766 \pm 0.006$ ps$^{-1}$ & \cite{HeavyFlavorAveragingGroupHFLAV:2024ctg} \\
\hline
$\mathcal{B}(\bar{B }\to X_s \gamma)$  & $(3.49\pm  0.19)\times10^{-4} $ & \cite{HeavyFlavorAveragingGroupHFLAV:2024ctg} \\
\hline\hline
\end{tabular}
\caption{Experimental inputs for the flavor observables used in this analysis.}
\label{tab:expt-data2}
\end{table}

\section{Results}\label{sec:results}
\begin{figure}[!ht]
\includegraphics[width=0.48\textwidth]{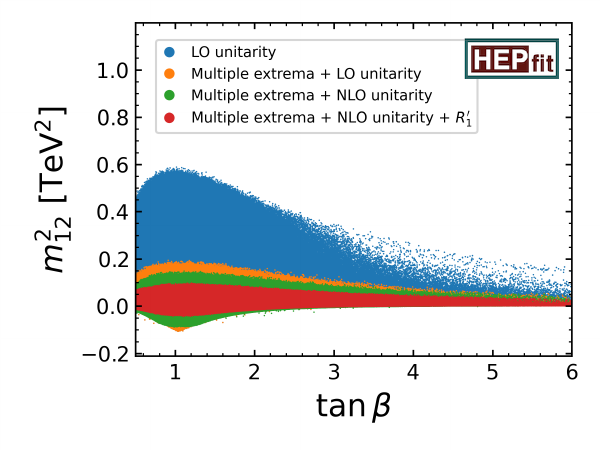}
	\caption{Allowed regions in the $\tan\beta $ vs.~$m_{12}^2$ plane consistent with the theoretical constraints. The blue points are allowed by the LO unitarity constraints. The orange, green, and red points satisfy the multiple-extrema condition after imposing the LO unitarity, NLO unitarity, and NLO perturbativity ($R_1^\prime<1$) conditions, respectively. }
    \label{fig:1}
\end{figure}

In this section, we present our results for the type-II 2HDM parameter space. In Figure~\ref{fig:1}, we compare limits on $m_{12}^2$ arising from the theoretical constraints mentioned above as a function of $\tan\beta$. The blue region denotes the allowed parameter space obtained by imposing LO unitarity conditions on the quartic couplings, while the orange region corresponds to the subset of the parameter space that may admit multiple extrema. Imposing NLO unitarity and perturbativity ($R_1' < 1$) further restricts this region, yielding the green and red regions, respectively. We observed that the unitarity conditions are satisfied for relatively large values of quartic couplings; however, the perturbativity restricts such large couplings, as reflected by the red region. The upper limit on $m_{12}^2$ in each case is a direct consequence of Eq.~\eqref{eq:upperbound_m12sq}, where the quantities $\Lambda_0$ and $\Lambda_1$ are bounded from above by the perturbativity and unitarity conditions, whereas the lower
limit on $m_{12}^2$ arises from the lower limit of the mass priors together with the perturbative unitarity conditions on the quartic couplings (see Figure~\ref{fig:A1} in Appendix~\ref{app:supp_plots}). The largest allowed magnitude of $m_{12}^2$, for both positive and negative values, occurs at $\tan\beta=1$, where the product $v_1 v_2$ is maximized. For large $\tan\beta$, $m_{12}^2 \geq 0$ due to the perturbativity constraint on the quartic coupling $|\lambda_1|<4\pi$. These constraints on $m_{12}^2$ translate into stringent bounds on $m_{H^\pm}$. The impact of these constraints in the $\tan\beta$ vs.~$m_{H^\pm}$ plane is shown in Figure~\ref{fig:A6} of Appendix~\ref{app:supp_plots}.

\begin{figure}[!ht]
\includegraphics[width=0.485\textwidth]{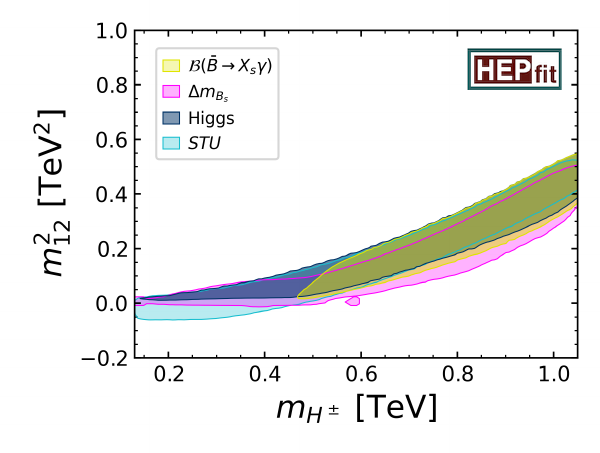}
	\caption{$95.4\%$ probability regions in the $m_{H^\pm} $ vs.~$m_{12}^2$ plane from experimental constraints. The yellow region represents the parameter space allowed from the $\mathcal{B}(\bar{B }\to X_s \gamma)$ data. The magenta and blue areas indicate the regions allowed from the $B_s-\bar{B}_s$ mixing data and the Higgs signal strengths data, respectively. The teal region is allowed by the $S$, $T$, and $U$ observables. }
    \label{fig:2}
\end{figure}

In Figure~\ref{fig:2}, we show the $95.4\%$ probability regions in the $m_{H^\pm}$ vs.~$m_{12}^2$ plane obtained from a fit to individual experimental constraints. In this fit, only perturbativity bounds ($|\lambda_i|<4\pi$) on the quartic couplings are imposed among the theoretical constraints. The constraint from $\mathcal{B}(\bar{B} \to X_s \gamma)$ induces a lower bound on $m_{H^\pm}$, which in turn translates into a lower bound on $m_{12}^2$, as shown in the yellow region. From the one-dimensional marginalized posterior, we find that $m_{H^\pm} \gtrsim 580~\mathrm{GeV}$ is allowed at $95.4\%$ probability by the latest $\mathcal{B}(\bar{B} \to X_s \gamma)$ data (see Figure~\ref{fig:6} in Appendix~\ref{app:supp_plots}).\footnote{Recently, the $\mathcal{O}(\alpha_s)$ corrections to $\mathcal{B}(\bar{B}\to X_s\gamma)$ have been improved by including four- and five-body final state contributions~\cite{Brune:2025zhd,Huber:2026pwu}. Ref.~\cite{Misiak:2026sqy} included these corrections in the Type-II 2HDM and obtained a stronger lower bound of $m_{H^\pm}>670$ GeV at 95\% C.L.  The combined global fit, which yields $m_{H^\pm} \gtrsim 725~\mathrm{GeV}$ (see Figure~\ref{fig:8} in Appendix~\ref{app:supp_plots}), does not yet include these new corrections. Including the updated $\mathcal{B}(\bar{B}\to X_s\gamma)$ constraint in the combined global fit is expected to further strengthen the lower bound on $m_{H^\pm}$, which we leave for future work.} This corresponds to a lower bound on $m_{12}^2 \gtrsim (283~\mathrm{GeV})^2$.

The blue region in Figure~\ref{fig:2} is allowed from the $h$ signal strength data. From the fit, we observe that the Higgs data also exclude the region where $m_{12}^2\leq 0$ at a $95.4\%$ probability. This is because, in the low-$m_{H^\pm}$ region, the charged scalar contribution interferes destructively with the $W$-boson contribution in the loop-induced $h \to \gamma\gamma, Z\gamma$ processes for $m_{12}^2\leq 0$, leading to a reduction in the respective signal strengths compared to the SM. On the other hand, in the large-$m_{H^\pm}$ region, $m_{12}^2$ should be positive to satisfy the perturbativity of the quartic couplings. Thus, the $Z_2$-symmetric limit $(m_{12}^2=0)$ is excluded in the Type-II 2HDM which indicates possible existence of multiple vacua, including vacua deeper than the EW vacuum, given the Higgs potential.

For completeness, we also consider EW precision and other relevant flavor observables in the fit. The magenta region in Figure~\ref{fig:2} shows the parameter space allowed by the $B_s-\bar{B}_s$ mixing data, while the allowed region after applying the EW precision constraints from the $S, T,$ and $U$ parameters is shown in teal. We find that neither the $B_s-\bar{B}_s$ mixing data nor the EW precision data significantly constrain the $m_{H^\pm} $ vs.~$m_{12}^2$ plane compared to the other two experimental observables. The corresponding results in the $\tan\beta $ vs.~$m_{12}^2$ plane are shown in Figure~\ref{fig:A2} of Appendix~\ref{app:supp_plots}.
 
\begin{figure}[!ht]
\includegraphics[width=0.485\textwidth]{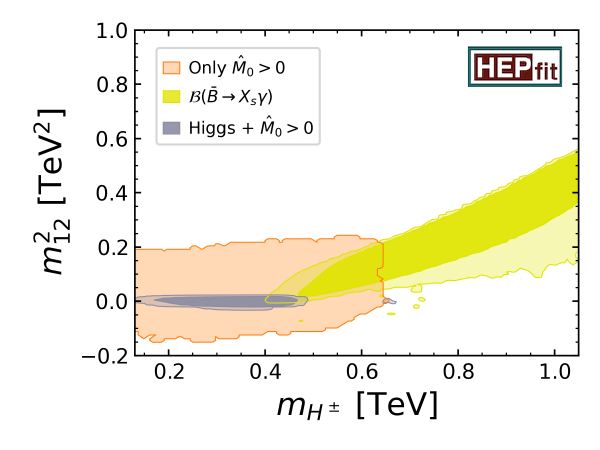}
	\caption{
    Allowed regions in the $m_{H^\pm} $ vs.~$m_{12}^2$ plane with tree level  unitarity and perturbativity constraints. The orange region in the parameter space satisfies the condition $\hat{M}_0>0$, while the gray region is allowed by the Higgs data with $\hat{M}_0>0$. The yellow region in the parameter space is allowed by the $\mathcal{B}(\bar{B} \to X_s \gamma)$ data. Dark (light) yellow/gray shaded area corresponds to 95.4\% (99.7\%) probability.}
    \label{fig:3}
\end{figure}

Figure~\ref{fig:3} shows the allowed parameter space where multiple extrema can occur in the $m_{H^\pm} $ vs.~$m_{12}^2$ plane. The orange region corresponds to the parameter space satisfying the multiple-extrema condition, i.e., $\hat{M}_0 > 0$, subject to the tree level unitarity and perturbativity constraints ($|\lambda_i|<4\pi$). Imposing the Higgs signal strength data substantially reduces this region to the gray area. This is because the LO unitarity condition rules out large scalar quartic couplings in the regime where $\beta-\alpha$ is close to $\pi/2$. Although Figure~\ref{fig:2} shows that the Higgs data exclude the region with $m_{12}^2 \leq 0$ at the 95.4\% probability level, some parameter points with $m_{12}^2 < 0$ reappear within the 95.4\% probability contour after imposing the theoretical condition $\hat{M}_0>0$. This occurs because the $\hat{M}_0>0$ condition removes theoretically disfavored regions from the fit, thereby reshaping the profile likelihood and the resulting probability contours.

The yellow region in Figure~\ref{fig:3} is allowed by the $\mathcal{B}(\bar{B }\to X_s \gamma)$ data without imposing the multiple-extrema condition ($\hat{M}_0>0$). The dark (light) yellow region corresponds to the $95.4\%$ ($99.7\%$) probability region. We find that the region satisfying both the multiple-extrema condition and the Higgs signal strength constraints does not overlap with the region favored by the flavor data at a $95.4\%$ probability (dark yellow). However, a small region of parameter space exists that can accommodate multiple extrema while satisfying both experimental constraints, i.e., Higgs signal strength and $99.7\%$ probability regions for flavor data (light yellow). It was shown in Refs.~\cite{Grinstein:2015rtl,Cacchio:2016qyh} that NLO unitarity more tightly constrains the quartic couplings than LO unitarity. Hence, in the following, we analyze the impact of the NLO unitarity  constraints on the parameter space that admits multiple extrema.

\begin{figure}[!ht]
\includegraphics[width=0.485\textwidth]{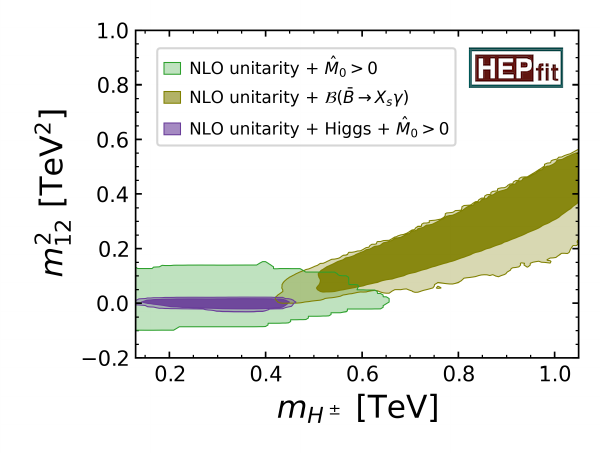}
	\caption{Allowed regions in the $m_{H^\pm} $ vs.~$m_{12}^2$ plane with NLO unitarity constraints. The olive region is allowed by the latest $\mathcal{B}(\bar{B }\to X_s \gamma)$ data in addition to the NLO unitarity conditions on the quartic couplings. The green region satisfies the condition $\hat{M}_0 > 0$ along with NLO unitarity bounds. The purple region represents the subset of the green region that also satisfies the Higgs signal strength data.  The dark (light) olive/purple areas correspond to 95.4\% (99.7\%) probability regions.}
    \label{fig:4}
\end{figure}

In Figure~\ref{fig:4}, the olive region denotes the parameter space consistent with the latest $\mathcal{B}(\bar{B }\to X_s \gamma)$ data when the NLO unitarity conditions are imposed on the scalar quartic couplings. The green region corresponds to the parameter space satisfying the multiple-extrema condition and the NLO unitarity constraints. This region is further reduced to the purple region once the Higgs signal strength data are included. The dark (light) olive/purple shaded areas correspond to the 95.4\% (99.7\%) probability regions. Similar to Figure~\ref{fig:3}, parameter spaces allowed by the Higgs signal strength and flavor data at a 95.4\% probability level (dark olive) do not overlap, after imposing the NLO unitarity bounds. However, while considering the 99.7\% probability-level regions (light olive) for the flavor data, a tiny overlapping region exists, but it is significantly reduced compared to the LO unitarity case (see Figure~\ref{fig:3}). This tiny overlapping region can accommodate multiple extrema while remaining consistent with the Higgs and flavor data. 

We  sample approximately $10^5$ parameter points within this overlapping region, each satisfying all the theoretical and experimental constraints discussed above. For each of these parameter points, we find that the physical EW vacuum is the global minimum of the Type-II 2HDM at tree level.\footnote{Recall that $\hat{M}_0>0$ is a necessary, but not sufficient, condition for the existence of multiple extrema.} Loop corrections can, in principle, modify the hierarchy among the vacua, such that a tree level global minimum becomes a local minimum at one loop, and vice versa. Such a situation can arise when two tree level minima are nearly degenerate, as demonstrated for the inert doublet model in Ref.~\cite{Ferreira:2015pfi}. To investigate this possibility, we use Eq.~(\ref{eq:hbar}) to evaluate the one-loop CW corrected potential at all neutral, CP-conserving tree level extrema and compare their depths. In all cases, we find that the physical EW vacuum remains the global minimum of the Type-II 2HDM even at the one-loop level.

\begin{figure}[!ht]
\includegraphics[width=0.485\textwidth]{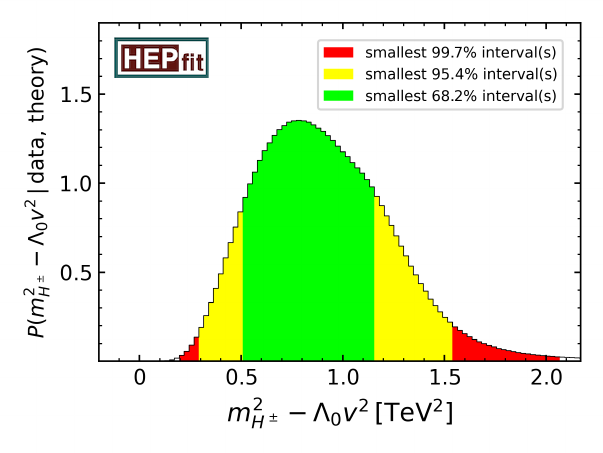}
	\caption{One-dimensional (1D) marginalized posterior probability distribution for $m_{H^\pm}^2-\Lambda_0 v^2$.}
    \label{fig:7}
\end{figure}

Combining all the above-mentioned experimental and theoretical constraints, including the NLO unitarity, we present in Figure~\ref{fig:7} the posterior distribution of $m_{H^\pm}^2-\Lambda_0 v^2$, marginalized over all other parameters. As discussed above,  the necessary condition for the tree level scalar potential to admit multiple stationary points (i.e., $\hat M_0>0$) can be recast in terms of physical parameters as $m_{H^\pm}^2-\Lambda_0 v^2<0$ (see Eq.~\eqref{eq:upperbound_mH+}). We find that the posterior probability for this condition, $P\!\left(m_{H^\pm}^2-\Lambda_0 v^2 < 0   \,\middle|\, \mathrm{data},\,\mathrm{theory}\right)$, is less than  0.003. Therefore, the parameter region corresponding to the existence of multiple vacua, and hence a metastable electroweak vacuum, is disfavored with a probability of more than $99.7\%$.

\begin{figure}[!ht]
\includegraphics[width=0.485\textwidth]{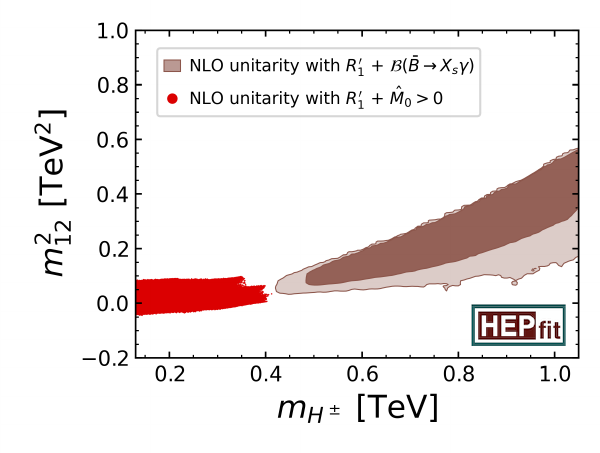}
	\caption{Allowed regions in the $m_{H^\pm} $ vs.~$m_{12}^2$ plane with NLO unitarity and $R_1' < 1$ constraints. The brown region denotes the allowed parameter space given the latest $\mathcal{B}(\bar{B }\to X_s \gamma)$ data with NLO unitarity + $R_1'<1$ conditions. Dark (light) brown areas correspond to 95.4\% (99.7\%) probability regions. The red regions have the same meaning as in Figure~\ref{fig:1}.  }
    \label{fig:5}
\end{figure}

In Figure~\ref{fig:5}, we illustrate the impact of the NLO unitarity constraints together with the perturbativity criterion ($R_1'<1$) on the global fit. The brown region denotes the parameter space allowed by the latest $\mathcal{B}(\bar{B}\to X_s\gamma)$ data after imposing the NLO unitarity and $R_1' < 1$, while the red region corresponds to parameter points that admit multiple extrema at the tree level. We observe that the region of parameter space accommodating multiple extrema is substantially reduced once the perturbativity condition is imposed, compared to the regions allowed by the LO or NLO unitarity constraints alone. This is because, although the LO or NLO unitarity bounds can allow large quartic couplings, the perturbativity of the series fails to hold for such large couplings. Consequently, the perturbativity criterion imposes stringent bounds on the parameter space that can accommodate multiple extrema, even before the Higgs signal strength constraints are taken into account. We find that no overlapping region of parameter space remains once the flavor constraints and the NLO unitarity, along with the perturbativity condition ($R_1' < 1$), are imposed. The same behavior is also observed in the $\tan\beta$ vs.~$m_{H^\pm}$ plane (see Figure~\ref{fig:A7} in Appendix~\ref{app:supp_plots}).

To summarize, the combined impact of perturbative unitarity, perturbativity, and the latest Higgs and flavor data leaves essentially no room for a metastable EW vacuum in the Type-II 2HDM.\footnote{In this work, we study the metastability of the EW vacuum only at zero temperature. At finite temperature, the Universe could in principle have become trapped in a different vacuum instead of the EW vacuum. Addressing this would require a finite temperature analysis of the effective potential and the corresponding vacuum-transition dynamics, which is beyond the scope of this work.} A deeper, coexisting minimum requires the scalar potential to admit multiple extrema, which in turn favors a relatively light charged Higgs boson, whereas the flavor data, in particular the $\mathcal{B}(\bar{B}\to X_s\gamma)$ bound, instead require a heavy one; tightening the theoretical constraints therefore drives the two regions apart. At the $95.4\%$ probability level, once the NLO unitarity is imposed, the region that can support multiple extrema no longer overlaps with the region preferred by the data. A small overlap persists at the $99.7\%$ probability level; there, however, the physical EW vacuum is found to be the global minimum at tree level, and, on evaluating the depths of the extrema with the $\hslash$ expansion, also at one-loop level. Furthermore, combining all the experimental and
theoretical constraints including NLO unitarity, we find the 1D marginalized posterior probability for the condition to admit multiple stationary points ($P\!\left(m_{H^\pm}^2-\Lambda_0 v^2 < 0   \,\middle|\, \mathrm{data},\,\mathrm{theory}\right)$) is less than 0.003. In addition, imposing the perturbativity criterion $R_1'<1$ removes this overlap entirely. We therefore conclude that, throughout the parameter space allowed by current data, the physical EW vacuum of the Type-II 2HDM remains the global minimum of the scalar potential at both tree and one-loop level, or that the possibility of a metastable physical EW vacuum is excluded.

\section{Conclusions}\label{sec:conclusions}
In this work, we analyze the vacuum structure of the Type-II 2HDM after imposing a comprehensive set of theoretical and experimental constraints. Because of the presence of two Higgs doublets in the 2HDM, in contrast to the SM, the physical EW vacuum need not correspond to the global minimum of the scalar potential and may instead be metastable, with a deeper minimum existing elsewhere in the field space.

We perform a global fit to the latest experimental data, together with the relevant theoretical constraints, including NLO unitarity, perturbativity, and BFB conditions. We find that the combined effect of the LO unitarity and Higgs data significantly reduces the parameter space that admits multiple extrema, whereas each constraint considered individually allows a substantially larger parameter space. The parameter space satisfying both the multiple-extrema condition and the Higgs signal strength data has no overlap with the region allowed by the flavor constraints at the 95.4\% probability level.

Going beyond tree level, NLO unitarity and perturbativity conditions ($R_1^\prime<1$) play a significant role in reshaping the allowed parameter space. Imposing the NLO unitarity bounds on the scalar quartic couplings further reduces the parameter space admitting multiple extrema compared to that allowed by the LO unitarity constraints. The parameter space satisfying the NLO unitarity constraints, together with the multiple-extrema condition and Higgs signal strength data, still does not overlap with the region allowed by the flavor constraints at the 95.4\% probability level, further strengthening the tension between these constraints. However, at the 99.7\% probability level, a tiny overlap region exists, within which we find that the physical EW vacuum is the global minimum at both the tree and one-loop levels. Furthermore, the 1D marginalized posterior probability for the condition allowing multiple stationary points is found to be less than 0.003 after imposing all experimental and theoretical constraints, including NLO unitarity.

Although the LO and NLO unitarity conditions can allow for sizable quartic couplings, perturbative expansion fails for such large couplings. The NLO perturbativity condition therefore imposes a stringent bound on the parameter space admitting multiple extrema, independent of the Higgs signal strength constraints (see the red region in Figure~\ref{fig:5}). This region has no overlap with the parameter space allowed by the flavor constraints at the 99.7\% probability level (see the brown region in Figure~\ref{fig:5}), thereby reinforcing the absolute stability of the physical EW vacuum.

\acknowledgments
We thank Mikolaj Misiak for helpful discussions regarding the updated $\mathcal{B}(\bar{B}\to X_s\gamma)$ constraint on the 2HDM. D.C.~and P.M.~acknowledge funding from the ANRF, Government of India, under grant ANRF/CRG/2021/007579. P.M.~also acknowledges the support from the
Department of Atomic Energy (DAE), Government of India, under Project Identification Number RTI 4002. The work of K.M.~was supported in part by the U.S.~National Science Foundation under Grant No.~PHY-2310497. S.S.~acknowledges funding from the MHRD, Government of India, under the Prime Minister’s Research Fellows (PMRF) Scheme. D.C.~also acknowledges support from an initiation grant IITK/PHY/2019413 at IIT Kanpur and funding from the Indian Space Research Organization (ISRO) under grant STC/PHY/2024427Q.

\appendix 
\section{Supplemental Plots}\label{app:supp_plots}
\begin{figure}[!ht]
\centering
\includegraphics[width=0.485\textwidth]{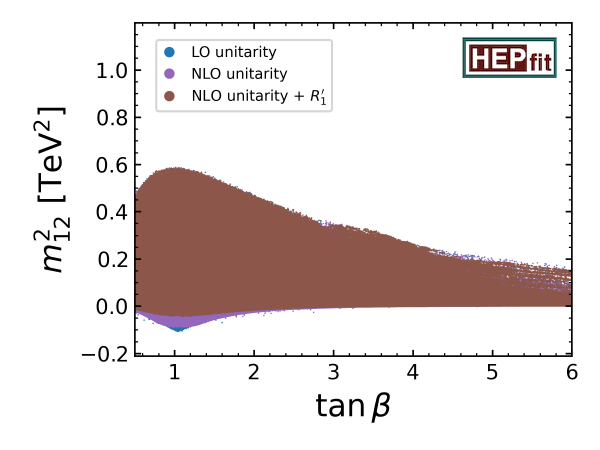}
	\caption{Allowed regions from the different perturbative unitarity constraints. The blue, purple, and brown points are allowed after imposing the LO unitarity, NLO unitarity, and NLO perturbativity ($R_1^\prime<1$) conditions, respectively. }
    \label{fig:A1}
\end{figure}

In this appendix, we present supplemental figures relevant to the analysis. In Figure~\ref{fig:A1}, we present the allowed parameter regions arising from the LO unitarity, NLO unitarity, and NLO perturbativity ($R_1^\prime<1$) conditions in blue, purple, and brown, respectively. We see that the lower bound on $m_{12}^2$ arises from the lower limit of the mass priors together with the perturbative unitarity conditions on the quartic couplings, whereas its upper bound is set by the upper limit of the chosen priors on the masses of the new scalars.

\begin{figure}[!ht]
\centering
\includegraphics[width=0.485\textwidth]{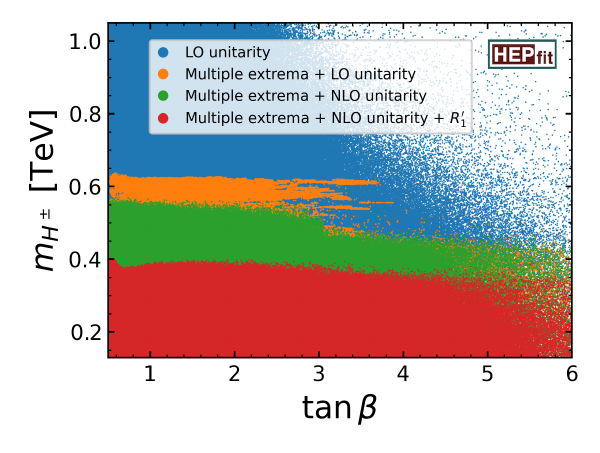}
	\caption{Allowed regions in the $\tan\beta$ vs. $m_{H^\pm}$ plane. The color regions have the same meaning as in Figure~\ref{fig:1}.}
    \label{fig:A6}
\end{figure}

Figure~\ref{fig:A6} illustrates how the upper bound on $m_{H^\pm}$ varies with $\tan\beta$ under different perturbative unitarity constraints, while satisfying the \em multiple-extrema\em ~condition $\hat{M}_0>0$. The color regions in Figure~\ref{fig:A6} have the same meaning as those in Figure~\ref{fig:1}. We find that the NLO unitarity constraint with the perturbativity condition $R_1' < 1$ places a stringent upper bound on the charged Higgs mass, $m_{H^\pm}\lesssim 400$ GeV, in the parameter space that allows multiple extrema.

\begin{figure}[!]
\includegraphics[width=0.485\textwidth]{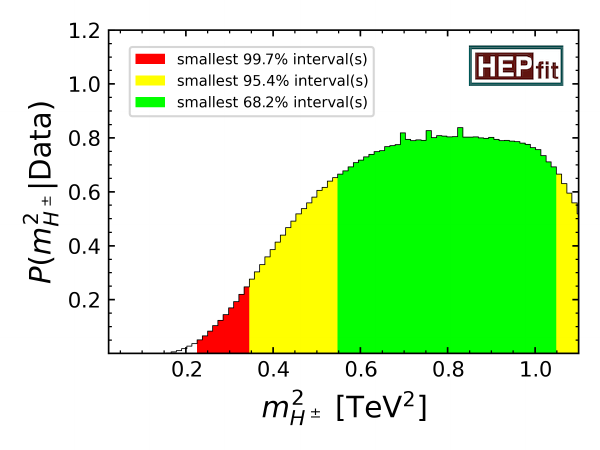}
	\caption{One-dimensional marginalized posterior probability distribution for $m_{H^\pm}^2$ derived from the latest $\mathcal{B}(\bar{B }\to X_s \gamma)$ data. The green, yellow, and red regions correspond to $68.2\% $, $95.4\%$, $99.7\%$ probabilities, respectively.}
       \label{fig:6}
\end{figure}

Figure~\ref{fig:6} shows the one-dimensional (1D) marginalized posterior probability distribution 
$P(m_{H^\pm}^2|\text{data})$ for $m_{H^\pm}^2$, with the smallest 68.2\%, 95.4\%, and 99.7\% credibility intervals indicated in green, yellow, and red, respectively. From the latest $\mathcal{B}(\bar{B }\to X_s \gamma)$ data, we find a lower bound on $m_{H^\pm}$ of approximately 580 GeV at a 95.4\% probability level.

\begin{figure}[!ht]
\centering
\includegraphics[width=0.485\textwidth]{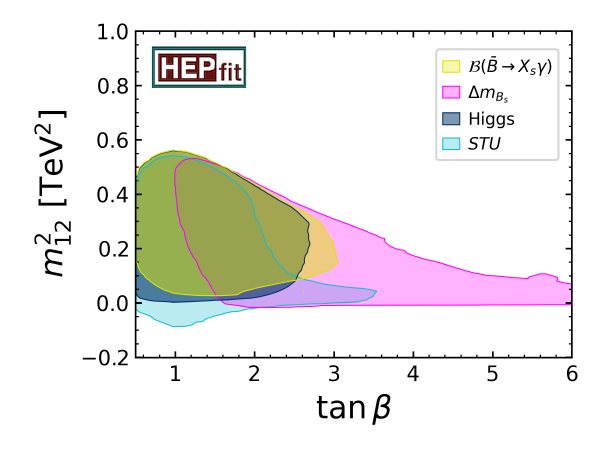}
	\caption{ $95.4\%$ probability regions in the $\tan\beta $ vs.~$m_{12}^2$ plane obtained from experimental constraints. The color regions have the same meaning as in Figure~\ref{fig:2}.}
    \label{fig:A2}
\end{figure}

In Figure~\ref{fig:A2}, we present the allowed regions in the $\tan\beta$ vs.~$m_{12}^2$ plane obtained from experimental constraints. Among the experimental constraints, $\mathcal{B}(\bar{B}\to X_s\gamma)$ provides a stringent lower bound on $m_{12}^2$ (see the yellow region). A stringent lower bound on $\tan\beta$ is obtained from the $\Delta m_{B_s}$ constraint, as shown by the magenta region.

\begin{figure}[!ht]
\centering
\includegraphics[width=0.485\textwidth]{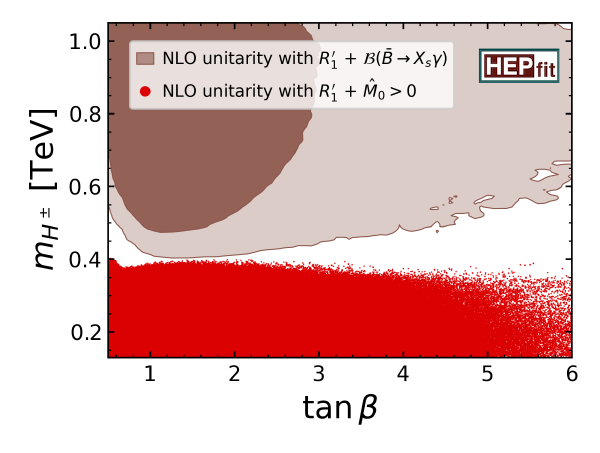}
	\caption{Allowed regions in the $\tan\beta$ vs. $m_{H^\pm}$ plane with NLO unitarity constraints. The color regions have the same meaning as in Figure~\ref{fig:5}.}
    \label{fig:A7}
\end{figure}

Figure~\ref{fig:A7} shows the allowed parameter regions in the $\tan\beta$ vs.~$m_{H^\pm}$ plane from the latest $\mathcal{B}(\bar{B }\to X_s \gamma)$ data with NLO unitarity + $R_1'<1$ conditions. The color regions in Figure~\ref{fig:A7} have the same meaning as those in Figure~\ref{fig:5}.  Similar to Figure~\ref{fig:5}, the region allowed by   the $\mathcal{B}(\bar{B}\to X_s\gamma)$ constraint at the $99.7\%$ probability level remains non-overlapping with the region admitting multiple extrema.

\begin{figure}[!htbp]
\includegraphics[width=0.485\textwidth]{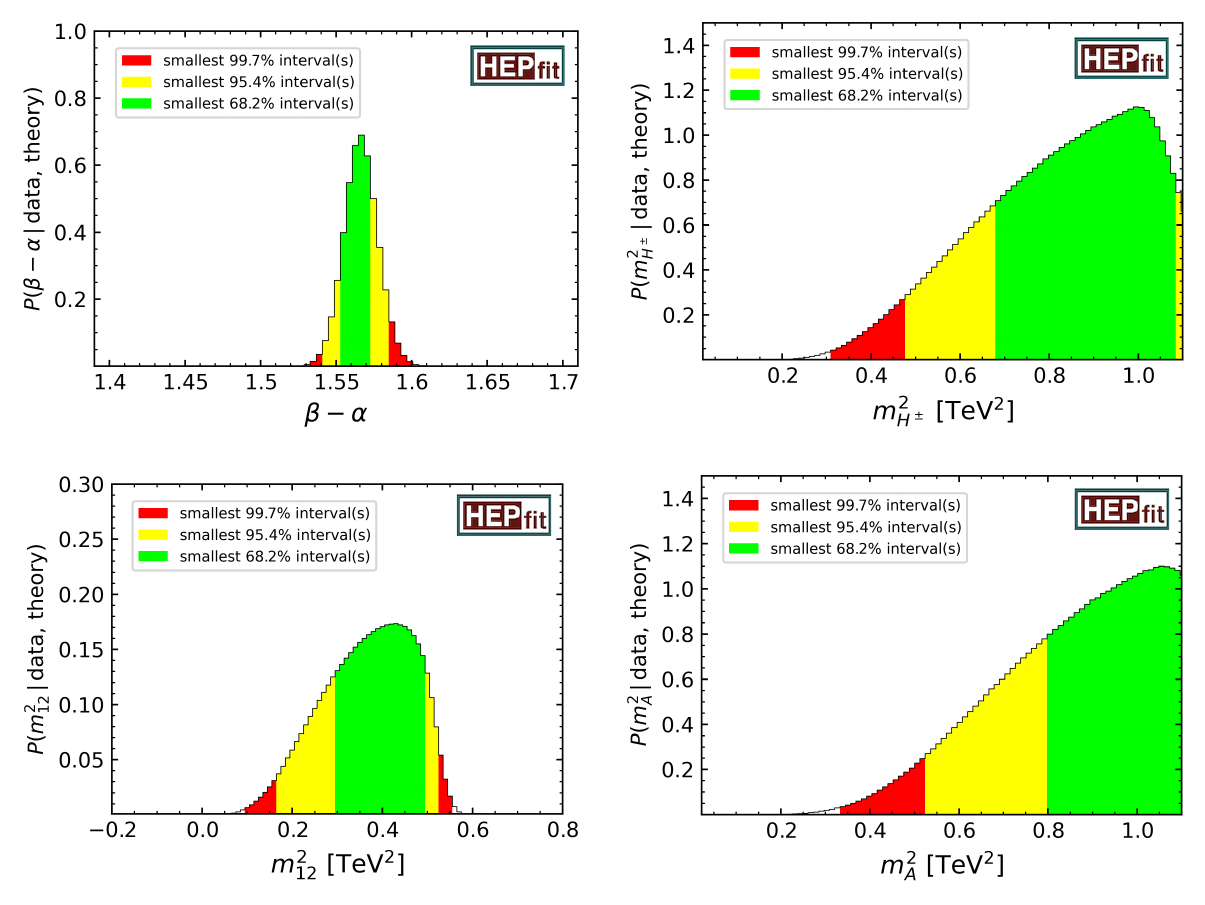}
	\caption{1D marginalized posterior probability distributions for 2HDM model parameters ($\beta - \alpha, m_{H^{+}}^{2}, m_{A}^{2},$ and $m_{12}^{2}$). The green, yellow, and red regions correspond to $68.2\% $, $95.4\%$, $99.7\%$ probabilities, respectively.}
       \label{fig:8}
\end{figure}

In Figure~\ref{fig:8}, we show the 1D marginalized posterior probability distributions for the model parameters, obtained by combining all the experimental and theoretical constraints, including the NLO unitarity. The green, yellow, and red regions correspond to the smallest $68.2\%$, $95.4\%$, and $99.7\%$ probability intervals, respectively. We find the following $95.4\%$ lower credible bounds: $m_{H^\pm}^2 \gtrsim 0.525~\mathrm{TeV}^2$, $m_{12}^2 \gtrsim 0.16~\mathrm{TeV}^2$, and $m_A^2 \gtrsim 0.515~\mathrm{TeV}^2$. The posterior distribution for
$\beta-\alpha$ is peaked around $1.564$, with the $95.4\%$ credible interval $1.539\lesssim \beta-\alpha \lesssim 1.587$\,.

\bibliographystyle{apsrev4-2}
\bibliography{main}
\end{document}